\documentclass[%
aip,
amsmath,amssymb,
reprint,%
]{revtex4-1}
\usepackage{dcolumn}% Align table columns on decimal point
\usepackage{bm}% bold math
\usepackage[utf8]{inputenc}
\usepackage[T1]{fontenc}
\usepackage{mathptmx}
\usepackage{textcomp}
\usepackage{amssymb}
\usepackage{amsmath}
\usepackage{bbm}
\usepackage{graphicx}
\usepackage{color}
\usepackage{ulem}
\usepackage{epstopdf}

\usepackage[colorlinks=true,citecolor=blue,urlcolor=blue]{hyperref}

\begin{document}
	
%	\preprint{AIP/123-QED}
	
\title{Shielded cantilever with on-chip interferometer circuit for THz scanning probe impedance microscopy }
%\author{M. Finkel, H. Thierschmann, M.P.Westig, A.J.Katan, M.Spirito and T.M. Klapwijk}
%\email[]{artanov@phystech.edu}
%\affiliation{Kavli Institute of NanoScience, Delft University of Technology, 
%Lorentzweg 1, 2628 CJ Delft, The Netherlands}
\author{Matvey~Finkel}
\affiliation{Kavli Institute of NanoScience, Department of Quantum Nanoscience, Faculty of Applied Sciences, Delft University of Technology, 
Lorentzweg 1, 2628 CJ Delft, The Netherlands}
\author{Holger~Thierschmann}
\email[]{h.r.thierschmann@tudelft.nl}
\affiliation{Kavli Institute of NanoScience, Department of Quantum Nanoscience, Faculty of Applied Sciences, Delft University of Technology, 
Lorentzweg 1, 2628 CJ Delft, The Netherlands}
\author{Allard~J.~Katan}
\affiliation{Kavli Institute of NanoScience, Department of Quantum Nanoscience, Faculty of Applied Sciences, Delft University of Technology, 
Lorentzweg 1, 2628 CJ Delft, The Netherlands}
\author{Marc~P.~Westig}
\affiliation{Kavli Institute of NanoScience, Department of Quantum Nanoscience, Faculty of Applied Sciences,  Delft University of Technology, 
Lorentzweg 1, 2628 CJ Delft, The Netherlands}
\author{Marco Spirito}
\affiliation{Department of Microelectronics, Faculty of Electrical Engineering,  Delft University of Technology, Mekelweg 4, 2629 JA Delft, The Netherlands} 
\author{Teun~M.~Klapwijk}
\email[]{t.m.klapwijk@tudelft.nl}
\affiliation{Kavli Institute of NanoScience, Department of Quantum Nanoscience, Faculty of Applied Sciences, Delft University of Technology, 
Lorentzweg 1, 2628 CJ Delft, The Netherlands}
\affiliation{Physics Department, Moscow State Pedagogical University, Moscow 119991, Russia.}
\date{\today}

\begin{abstract}

We have realized a microstrip based THz near field cantilever which enables quantitative measurements of the impedance of the probe tip at THz frequencies (0.3 THz). A key feature is the on-chip balanced hybrid coupler which serves as an interferometer for passive signal cancellation to increase the readout circuit sensitivity despite extreme impedance mismatch at the tip. We observe distinct changes in the reflection coefficient of the tip when brought into contact with different dielectric (Si, SrTiO$_3$) and metallic samples (Au). By comparing to finite element simulations we determine the sensitivity of our THz probe to be well below $ 0.25~$fF. The cantilever further allows for topography imaging in a conventional atomic force microscope mode. Our THz cantilever removes several critical technology challenges and thus enables a shielded cantilever based THz near field microscope.

%This allows us to  We launch a THz signal (225-325 GHz) to the transmission line  which enables to perform reflection measurements on the cantilever tip. At one end of the cantilever the THz circuitry terminates in a metallic tip serving as a near field probe. At the other end two ports, compatible with standard landing probes allow for injecting and detecting a THz signal.  The two ports and the metallic tip are connected through a branchline hybrid coupler which allows to measure the reflection at the tip by monitoring the transmission between ports 1 and 4.
%
%on the other end. Our circuitry enables measurement of a THz signal which is launched on the circuitry and reflected at the cantilever tip. This allows us to detect a change of the tip impedance due to a dielelectric material which is braought into contact with the tip. Combining our cantilever with a scanning probe setup will provide a major step towards the scanning THz nearfield impedance microscope.
%
\end{abstract}
\maketitle

\section{Introduction}
In condensed matter research there is a strong desire for new experimental tools that can measure the local electrical properties of materials and buried layers at high frequencies and with high spatial resolution  \cite{basov2011electrodynamics}.
For this goal, in recent years a variety of powerful scanning probe techniques have emerged which cover different parts of the electromagnetic spectrum: on the one hand, scanning near-field optical microscopy (SNOM) has enabled imaging with %optical \cite{gross2018near}, 
infrared and far-infrared \cite{qazilbash2007mott,liu2016nanoscale,huber2008terahertz} frequencies down to a few THz by utilizing the optical toolbox, i.e. free-space radiation, lasers and fiber technology; on the other hand, at GHz frequencies (typically 1-20 GHz) co-axial probes and shielded cantilevers have made possible quantitative local imaging by making use of commercially available microwave electronics for high performance signal processing (scanning microwave impedance microscopy, SMIM) \cite{lai2010mesoscopic,de2016spatial,gramse2017nondestructive,anlage2007principles,steinhauer1998quantitative,buchter2018scanning,tselev2007broadband,huber2012calibrated}.  
The frequency band in between, however, ranging approximately from 100 GHz to a few THz (also referred to as \textit{sub-mm waves}), is a technological challenge. In the field of astronomy detection major progress has been made in sub-mm technology, for instance in the development of phase preserving instruments based on superconducting tunnel junctions such as for the Herschel Space Telescope, the Atacama Pathfinder Experiment (APEX) and the Atacama Large Millimeter array (ALMA)\cite{eht2019black,siegel2002terahertz}.
These advances are being picked up to promote technological progress also in other research fields. In condensed matter physics this is expected to have a strong impact on measurement instrument development which will help understanding of a variety of important problems, in particular for disordered and unconventional superconductors, as well as for the so-called quantum materials where strong electron-electron interactions in the THz energy range give rise to a number of puzzling, unconventional and often spatially inhomogeneous electrical properties \cite{liu2016nanoscale,basov2011electrodynamics,dordevic2006electrodynamics}. 
%Having experimental tools available to locally probe these material's electrodynamic behavior at THz frequencies is thus highly desirable.    
Realizing an experimental tool for probing these properties, however, remains challenging. Only recently first scattering SNOM measurements below 1 THz have been reported \cite{liewald2018all}. 
In the sub-mm band on-chip electronic circuits suffer from high losses (at room temperature) which strongly complicates the fabrication of more complex circuitry, required for signal control and processing. An alternative technology, much less prone to losses, consists of quasi-optical and metallic waveguide components. However, this technology increases the size of the measurement instrument compared with on-chip circuitry and thus also imposes certain boundaries when more complex signal handling is needed.
In order to overcome these technological hurdles, there is an ongoing development to combine quasi optical and on-chip electronics in hybrid devices \cite{westig2011balanced}, but also to push the performance of microwave electronics into the sub-mm-band \cite{zmuidzinas2004superconducting}. Picking up on this development, we have recently reported on a microstrip (MS) fabrication technology based on PECVD SiN$_x$, that is compatible with thin film membranes. For this technology losses are sufficiently well controlled at frequencies around 0.3 THz, such that the realization of room-temperature THz on-chip components is feasible \cite{finkel2017performance}. 
Here, we use this technology to extend scanning impedance microscopy from microwaves into the THz frequency range. We present a shielded THz cantilever suitable for scanning probe microscopy that enables quantitative measurement of the impedance of the cantilever tip at around 0.3 THz. A key ingredient is a branchline coupler which is patterned on the cantilever and which acts as an interferometer for the THz signal, thereby providing high sensitivity of the circuit to small impedance changes at the cantilever tip. 

First, we will revisit the concept of scanning near field microscopy with shielded cantilevers as it is currently being used in microwave microscopy and we will identify the key features a THz cantilever should comprise. We then present the concept of our THz on-chip interferometer. Finally, we demonstrate how this concept enables impedance measurements with a THz cantilever that is compatible with conventional atomic force microscopy (AFM).

\section{Principle of shielded cantilever microscopy}

\begin{figure}
      	\includegraphics[width=1\linewidth]{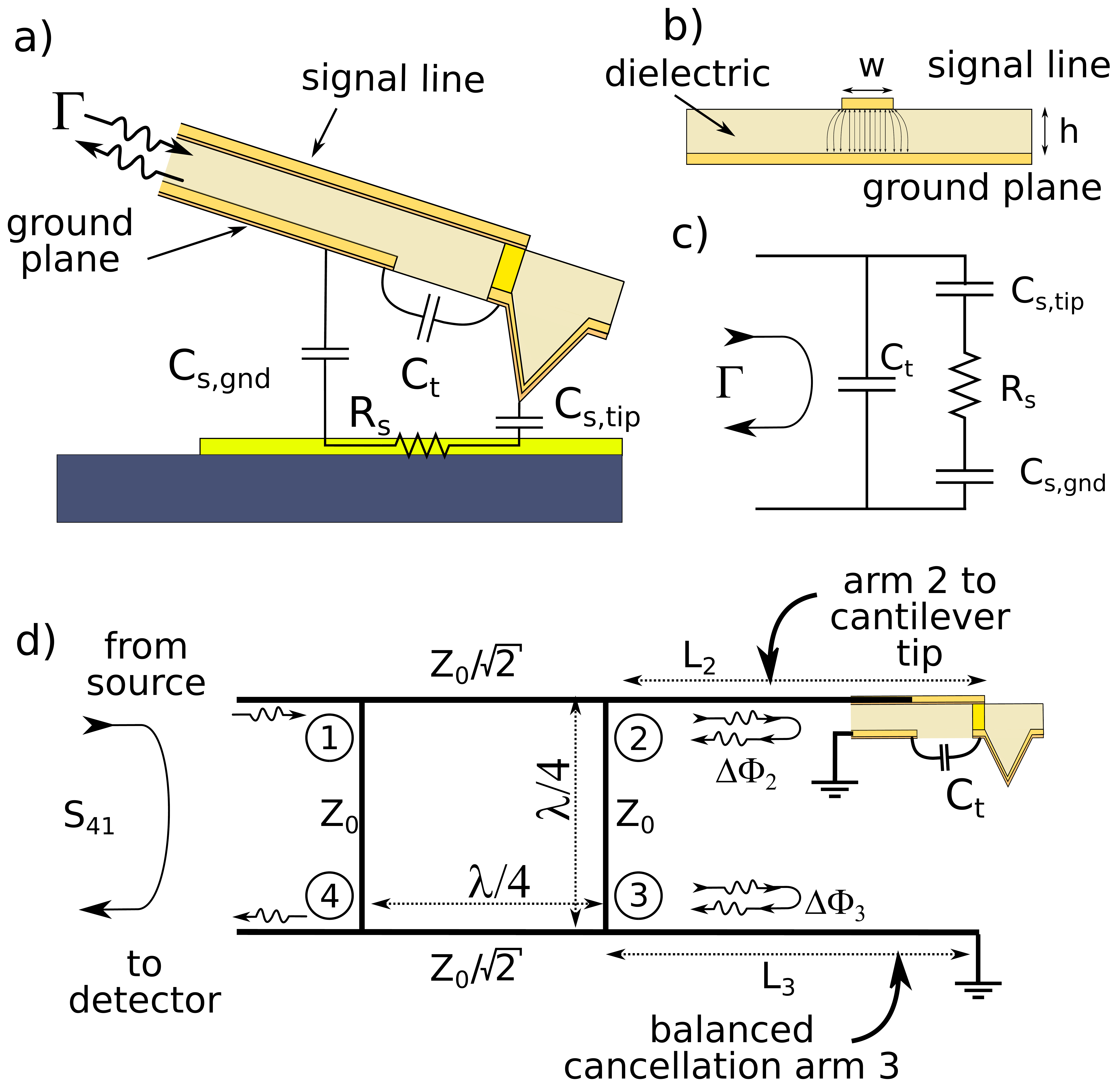}
      	\caption{\textbf{a)} Sketch of a shielded cantilever and the corresponding lumped element circuit for measuring the complex reflection coefficient $\Gamma$ of the cantilever tip. The cantilever consists of a metallic signal line and groundplane, separated by a dielectric, which determines the mechanical properties of the cantilever. The probe-cantilever interaction can be described with a lumped element circuit.  \textbf{b)} Cross section of the microstrip shielded cantilever. $w$ and $h$ denote the width of the signal line and the height of dielectric layer, respectively. \textbf{c)} Equivalent lumped element circtuit of the situation depicted in a). \textbf{d)} Distributed element circuit diagram for measuring the cantilever impedance using a balanced branchline coupler. The reflection coefficient $\Gamma$ at the cantilever tip is determined by measuring the scattering parameter $S_{41}$ of the branchline coupler. This is achieved by intereference of the reflected signal from the tip with an unknown phase shift $\Delta \Phi_2$ with that of a balanced cancellation arm with known phase shift $\Delta \Phi_{3}$. $Z_0$ denotes the characteristic line impedance and $\lambda$ is the signal wavelength. }
      	\label{fig:principle}
      \end{figure}

 The principle of scanning near field microscopy with a shielded microstrip (MS) cantilever is illustrated in Fig.~\ref{fig:principle}a) \cite{anlage2007principles,lai2011nanoscale}. The cantilever consists of a dielectric membrane of which the bottom side is covered with a thin metal layer, serving as a transmission line ground plane. 
 The signal line of width \textit{w} is patterned on the top side of the same dielectric. A cross section of the resulting MS transmission line geometry is sketched in Fig.~\ref{fig:principle}b). Because the high frequency fields are mostly confined within the dielectric, such a transmission line geometry allows for delivering the signal to the cantilever tip in a controlled way while the ground plane screens the environment and prevents radiation losses. At the end of the cantilever the signal line terminates in a metallic tip. When a high frequency tone is launched to the MS, the tip acts as a capacitive termination, reflecting the signal back into the cantilever. This is quantified by the reflection coefficient $\Gamma$, which is given by the mismatch between the generally complex valued tip impedance $Z$ and the characteristic MS line impedance $Z_0 = 50~\Omega$: $\Gamma={(Z-Z_0)}/{(Z+Z_0)}$. When the cantilever is lifted far away from the sample surface, $Z$ is given by the capacitance $C_t$ between the tip and the cantilever ground plane (see Fig. \ref{fig:principle} a). When the tip is on a sample, $Z$ is modified by contributions from the tip-sample capacitance $C_{s,tip}$, the capacitance between the sample and the ground plane $C_{s,gnd}$ and from resistive losses inside the sample, $R_s$. Measuring changes in $\Gamma$ by detecting phase and amplitude of the reflected signal while scanning the tip over the sample provides a quantitative image of the spatial distribution of the conducting and dielectric properties of the sample \cite{lai2010mesoscopic}. Since the electric field becomes strongly enhanced at the sharp tip, these local contributions dominate the total response, thus enabling spatial resolution down to 100 nm, i.e. three orders of magnitude below the signal wavelength \cite{,lai2010mesoscopic,gramse2017nondestructive}.  
 
 The tip-ground plane capacitance C$_t$ is generally given by the size and geometry of the cantilever close to the tip, which results in values of the order of $C_t \sim 10^{-15}F$ . Since $Z\propto 1/i\omega C$, at GHz frequencies ($\omega = 2\pi f$, $f \sim 10^9$ Hz) the terminating impedance is large $Z \sim 10^6 \Omega$ $ \gg Z_0$ and therefore the reflection coefficient becomes $\Gamma \simeq 1$. This means that most of the signal is reflected back into the cantilever when the tip is floating over the sample. We will refer to this part of the signal as scattered signal because it does not carry information about the sample itself. When the tip is in contact with the sample, the desired contributions from the tip-sample interaction thus only lead to small variations on top of an otherwise large $\Gamma$, which is obviously difficult to detect. It is therefore highly desirable to minimize the scattered signal in the detector line and to become sensitive to those contributions only, which originate from the tip-sample interaction. At microwave frequencies this problem has been addressed by making the cantilever and the tip part of a resonator \cite{cui2016quartz,gramse2017nondestructive,huber2012calibrated,anlage2007principles,tselev2007broadband} or by adding an impedance matching circuit which matches the open tip impedance to $Z_0$ \cite{lai2010mesoscopic}. Both solutions create a narrow band resonance condition which enhances the sensitivity of the circuit to changes in the tip impedance. Furthermore, a common mode cancellation loop is typically included into the microwave readout circuit \cite{huber2012calibrated,lai2010mesoscopic} which further reduces the scattered signal level at the detector.
 
 While at microwave frequencies such circuitry can be incorporated rather easily, this is not straight forward at THz frequencies because many required technologies are not readily available. In order to realize a shielded impedance microscope cantilever at THz frequencies it is therefore plausible to aim for an on-chip THz circuit solution that can be patterned close to the tip with lithographical means. We identify the following key properties such a circuit should provide: 1) separation of the in-going and reflected signal to facilitate signal processing. 2) Cancellation of the scattered signal in the detector line. 3) Sensitive response when the tip is brought into contact with a dielectric or a metallic sample. 4) Short signal lines to minimize losses. 
 In the following section we will present and demonstrate a circuit that fulfills all of these requirements.

\section{Balanced branchline coupler as on-chip interferometer} 
Figure \ref{fig:principle}d) depicts the diagram of a circuit designed to accomplish the above criteria. 
A key component is the balanced branchline coupler. It consists of 4 ports (labelled 1 - 4) which are connected through transmission line segments of a quarter wavelength $\lambda/4$ of the aimed for measurement frequency. By properly designing the impedance of each branch of the coupler (i.e. by choosing the appropriate signal line width $w$ for a constant thickness of the dielectric layer, cf. Fig.\ref{fig:principle}b), one can control the transmission coefficients between the ports.      
 The key idea of the concept we introduce here derives from analogies between a branchline coupler and an optical beam splitter: When the branch impedances $Z$ are chosen such that for two opposite branches $Z=Z_0$ ($w = 3.75~\mu$m), while for the other two $Z = Z_0/\sqrt{2}$ ($w = 7.5~\mu$m), an incoming signal at, for example, port 1, is split in equal parts between ports 2 and 3, and it acquires an additional phase shift of $-\pi/2$ between the these ports, while no signal arrives at port 4 \cite{pozar2009microwave}. Since the coupler is designed symmetrically, the signal is split in the same fashion when injected at any other port.
 %Note however, that the signal at port 2 

 We can now use these properties, signal splitting and phase delay, to build an on-chip interferometer that is highly sensitive to impedance changes at the cantilever tip: We attach transmission lines of finite length $L_2$ and $L_3 = L_2 + \Delta L$ at ports 2 and 3, respectively, as shown in Fig. \ref{fig:principle}d (which we will refer to as \textit{arms}, in analogy to an optical Michelson-interferometer). As the signal gets reflected at the end of each arm, it picks up a phase shift and gets re-injected into the coupler. For simplicity, assuming an ideal coupler with perfect isolation \cite{pozar2009microwave} and neglecting losses, the signal at port 4 (detector line) is then given by the sum of the reflected signals re-injected at port 2 and 3,
 
 \begin{equation}
 	S_{41} = \frac{A}{2} e^{i (\Phi_c + \Phi_{L2} + \Delta \Phi_2)} +  \frac{A}{2}  e^{i (\Phi_c + \Phi_{L3} +\Delta \Phi_3)},
 	\label{eq:coupler} 
 \end{equation}
 where $A$ corresponds to the total signal amplitude, $\Phi_c = 3\pi/2$ is the total phase accumulated in the coupler, $\Phi _{L2,3}$ refer to the phase picked up due to the signal traveling down the respective arms and $\Delta \Phi_{2,3}$ is the phase picked up due to reflection at the terminations of arms 2 and 3, respectively. For our purposes it is convenient to express Eq.\ref{eq:coupler} as 
 
  \begin{equation}
 S_{41} = \frac{A}{2} e^{i (\Phi_c + \Phi_{L2})} (e^{i \Delta \Phi_2} + e^{i (\Delta \Phi_3 + \Phi_{\Delta L})}).
 \label{eq:coupler_DeltaLeq0} 
 \end{equation}
 
 which indicates that signal cancellation in the detector line is achieved for 
 \begin{equation}
 \Delta \Phi_2 = \Delta \Phi_{3} + \Phi_{\Delta L} - \pi .
 \label{eq:phase}
 \end{equation}

 As shown in Fig.~\ref{fig:principle}d), in our case arm 2 terminates in the cantilever tip, which, in a first approximation  ($C_t \rightarrow 0$), acts as an \textit{open} termination ($\Delta \Phi_2 \rightarrow -\pi$) when the tip is lifted off the sample. It is therefore convenient to terminate arm 3 with a \textit{short} ($\Delta \Phi_3 = 0$) and to choose $\Phi_{\Delta L} = 0$ to achieve good signal cancellation at port 4. 
 When scanning, changes in the dielectric or metallic environment of the tip lead to a phase mismatch at the detector line due to an enhanced capacitance at the tip, according to Fig.~\ref{fig:principle}c. This results in a measurable signal which can be directly related to the phase change due to modified reflection conditions at the tip, using Eqs. \ref{eq:coupler_DeltaLeq0} and \ref{eq:phase}. When the tip is landed on a fully metallic sample, $(1/C_{s,tip} + 1/C_{s,gnd})^{-1} \gg C_t$. As can be seen from the circuit in Fig.~\ref{fig:principle}c) this corresponds to arm 2 being effectively shorted. In this case the reflected signals will interfere constructively and the full signal is detected at port 4.   
  We note that in a real cantilever $C_t$ is finite ($\Delta \Phi_2 \gtrsim -\pi$) in which case $\Delta L$ can be used as an additional phase matching parameter to achieve cancellation of the scattered signal.     

\begin{figure}
	\includegraphics[width=0.7\linewidth]{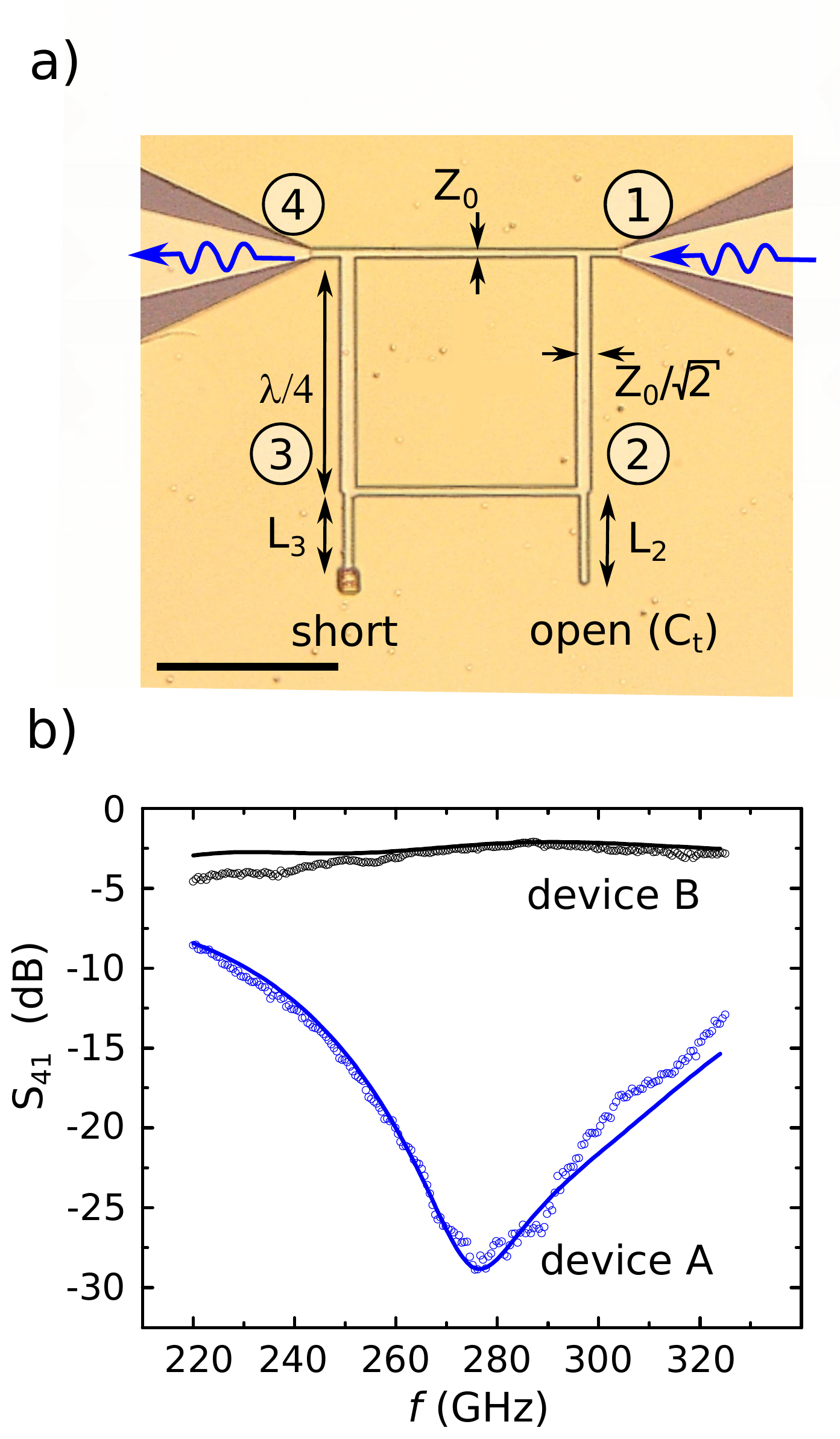}
	\caption{\textbf{a)} Optical image of the on-chip interferometer (device A), containing a balanced branchline coupler with branch lengths $\lambda/4$. The ports of the coupler are denoted 1-4. At a distance $L_2$ and $L_3$ from port 2 and 3 the transmission lines terminate in an open and short circuit, respectively. The signal (indicated with blue arrows) is injected at port 1 and detected at port 4. The scale bar corresponds to $100~\mu$m. \textbf{b)} S-parameter magnitude of a transmission measurement (symbols) and the corresponding analytic calculation (solid line) of device A (blue) and B (black). $L_2 = L_3 = 49 ~\mu$m.}
	\label{fig:branchline}
\end{figure}
 
 In order to test this concept, we have realized a series of balanced branchline couplers on a Si substrate using the technology described by Finkel \textit{et al.}\cite{finkel2017performance} All structures consist of $3~\mu$m thin SiN$_x$ ($\epsilon_r = 5.9$) serving as a MS dielectric and 2/300 nm Ti/Au as ground plane and stripline. As THz source and detector we use a vector network analyser together with frequency multipliers that cover the WR-03 band (220 to 325 GHz) and a GSG landing probe setup (for details see Finkel et al, Ref.\cite{finkel2017performance}). Note however, that our concept is also compatible with other THz sources and detectors, for instance photomixers \cite{mayorga2012first,westig2019waveguide}.  We first demonstrate conceptually the basic idea. For this we have fabricated two samples (device A and B) for which $\Delta L = 0$ and which realize two different arm terminations \textit{open/short} and \textit{open/open} at ports 2 and 3, respectively. An optical image of device A is shown in Fig.~\ref{fig:branchline}a. The signal is launched and picked up from the circuit via the ports labeled 1 and 4 in Fig.~\ref{fig:branchline}a, which consist of co-planar waveguide type fixtures\cite{finkel2017performance} (not visible) that enable coupling of THz signals into the circuit with the landing probes. 
 The $\lambda/4$-branches of the coupler have a length of 130 $\mu$m, corresponding to a branchline coupler center frequency of $f_c = 270$ GHz. For the arm lengths we choose $L_2= L_3 = 49~\mu$m.

Figure \ref{fig:branchline} b shows the measured scattering parameter $S_{41}$ obtained for device A and B (blue and black symbols, respectively). As expected, for device A we observe a low transmission ($\sim -30$~dB) between port 1 and 4 with a minimum at $f =280$ GHz, which is close to the branchline coupler's center frequency $f_c = 270$ GHz. For device B both arms terminate in an open, i.e. $\Delta \Phi_2 = \Delta \Phi_3$. As a result constructive interference leads to a high transmission ($\sim -5$~dB) over the full frequency range. 

\begin{figure}
	\includegraphics[width=0.6\linewidth]{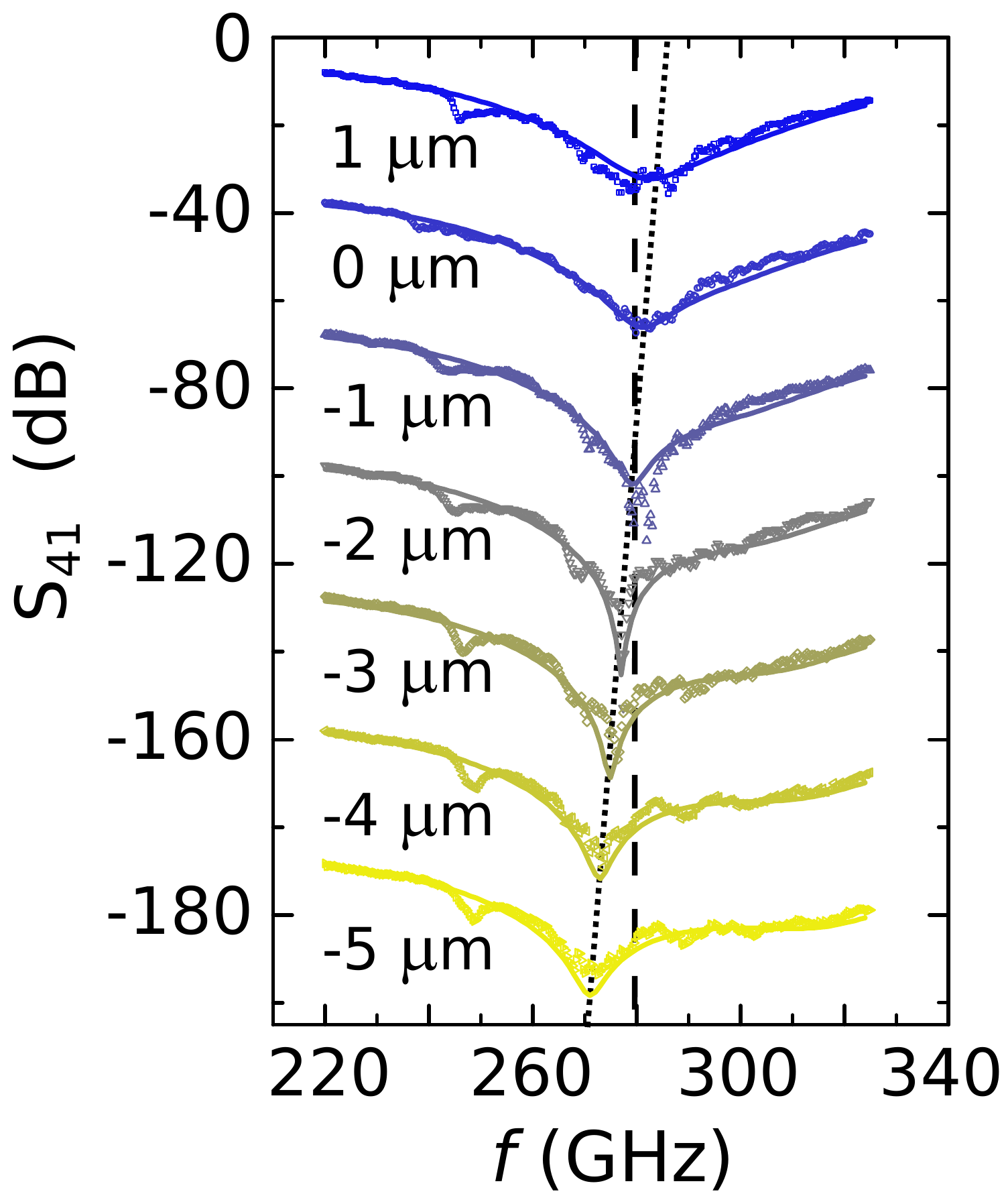}
	\caption{ Transmission measurements (symbols) and calculation (solid line) for a set of \textit{open/short} circuits with $L_3 = L_2 + \Delta L$ and $\Delta L = +1,0,...,-5~\mu$m and $L_2 = 49~ \mu$m. The curves are offset by -30 dB for clarity. }
	\label{fig:varyL}
\end{figure}

\begin{figure}
	\includegraphics[width=0.8\linewidth]{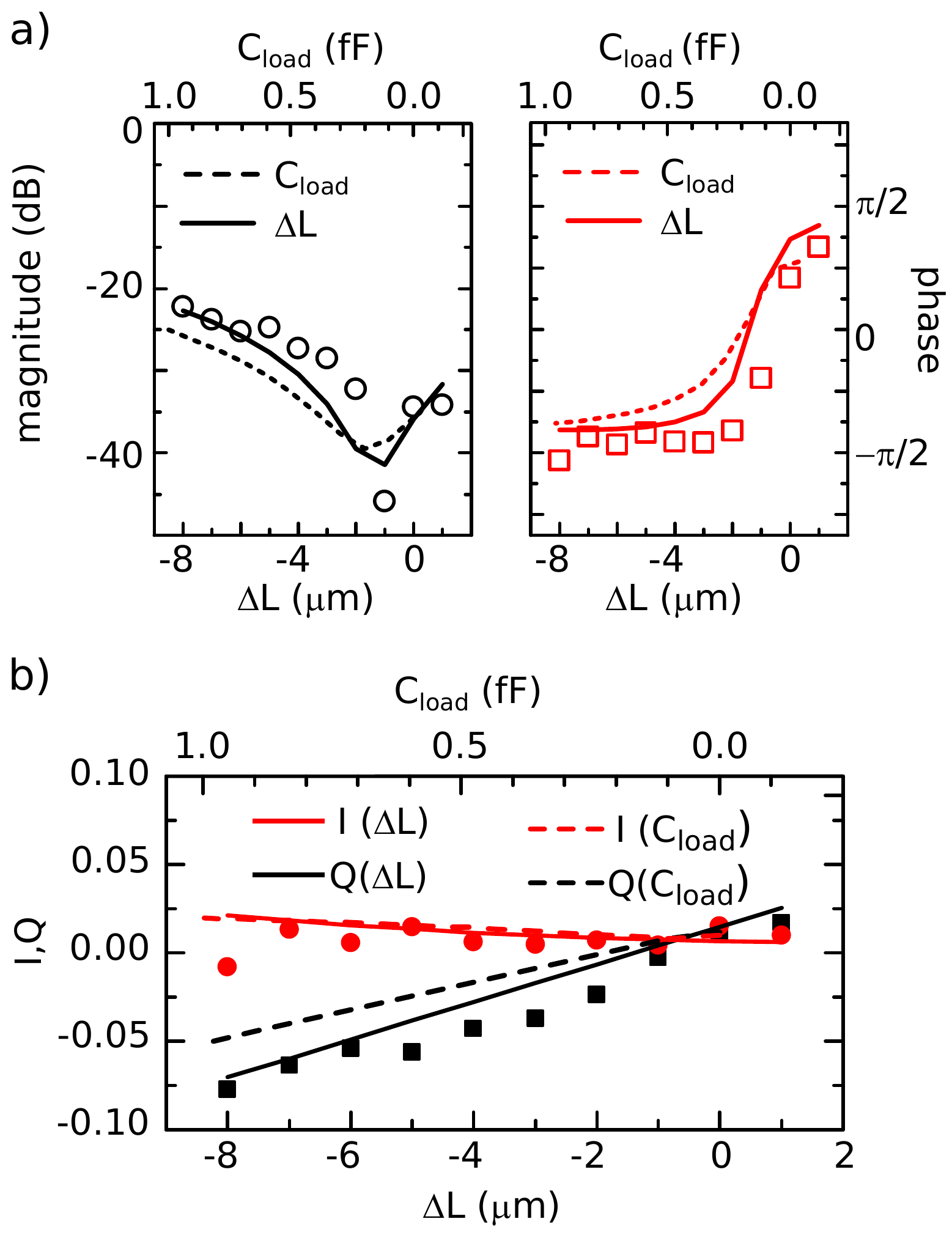}
	\caption{\textbf{a)} Magnitude (left) and phase (right) as function of $\Delta L$ (bottom axis). Symbols: measurements. solid line: calculation. Dashed line and top axis: Calculated phase and magnitude for $L_2=L_3 = 49~\mu$m and with varying $C_{load}$ connected in parallel to the termination of interferometer arm 2. \textbf{b)} In-phase (I, red) and quadrature (Q, black) representation of the measured (circles and squares) and calculated (solid and dashed line) response versus $\Delta L$.  Dashed lines and top axis: Calculated I (red) and Q (black) as a function of $C_\text{load}.$}
	\label{fig:Hyb_IQ}
\end{figure}

Next we demonstrate how, owing to the sharp interferometer cancellation conditions, the circuit is highly sensitive to contributions from $\Phi_{\Delta L}$. Figure \ref{fig:varyL} shows the measured $S_{41}$ parameter obtained from a series of devices for which we have varied $L_\text{3} = 49~\mu$m$ + \Delta L$ by $\Delta L = (1, 0, -1...-5)~ \mu$m, while leaving $L_\text{2} = 49~\mu$m fixed. This leads to a small phase imbalance for the signal paths along arms 2 and 3. The experimental data reveal that indeed the position of the dip in frequency as well as its depth sensitively depend on $\Delta L$ (dotted line). In Fig.~\ref{fig:Hyb_IQ}a we have extracted magnitude and phase (symbols) for each $\Delta L$ at fixed frequency $f = 280$ GHz (dashed line in Fig.\ref{fig:varyL}). The data show that signal cancellation improves for small $\Delta L$ with an optimal configuration at $\Delta L = -1~\mu m$. For even larger length difference it levels off. As discussed above this behavior reflects the termination of arm 2 with a finite capacitance, leading to phase shift slightly different from $-\pi$, which gets compensated for by a slightly shorter $L_3$.
This has been confirmed quantitatively within a textbook analytical model of the circuit \cite{pozar2009microwave} (for details see Appendix and Supplementary Material) that nicely reproduces all of our experimental data consistently (solid lines in Fig.~\ref{fig:varyL} and Fig.~\ref{fig:Hyb_IQ}a). In addition to a small dissipative contribution in the via, $R_\text{short} = 1.6~\Omega$, we have taken into account a finite terminating capacitance $C_{t} = 0.163$~fF, consistent with a standard text book approximation for an open MS line (see Appendix).

We can further use the analytical model to analyse theoretically the circuit's response to a load capacitance $C_\text{load}$ connected in parallel to $C_t$, representing a sample in a scanning probe experiment (cf. lumped element diagram in Fig. \ref{fig:principle}c). The resulting amplitude and phase are plotted in Fig.~\ref{fig:Hyb_IQ}a) as dashed lines. The corresponding $C_\text{load}$ is given in the top axis. As expected this yields a fairly similar behaviour as a variation of $\Delta L$. 
%This confirms the suitability of our circuit for sensitive impedance measurements. 
Figure ~\ref{fig:Hyb_IQ}b) plots the data as in-phase (I) and quadrature (Q) amplitudes, for a variation of $\Delta L$ (bottom axis, solid lines) and $C_\text{load}$ (top axis, dashed lines), respectively. In this representation $I$ can be directly related to dissipative contributions to the signal while $Q$ represents the imaginary part of the reflection coefficient which is related to capacitive (and, in principle, also inductive) contributions. This is consistent with the observed linear behaviour of $Q$ and a constant $I$. From these plots we estimate our circuit to be sensitive to a capacitance change smaller than 0.25 fF.

 \section{Cantilever implementation}
 
\begin{figure}
 	\includegraphics[width=1\linewidth]{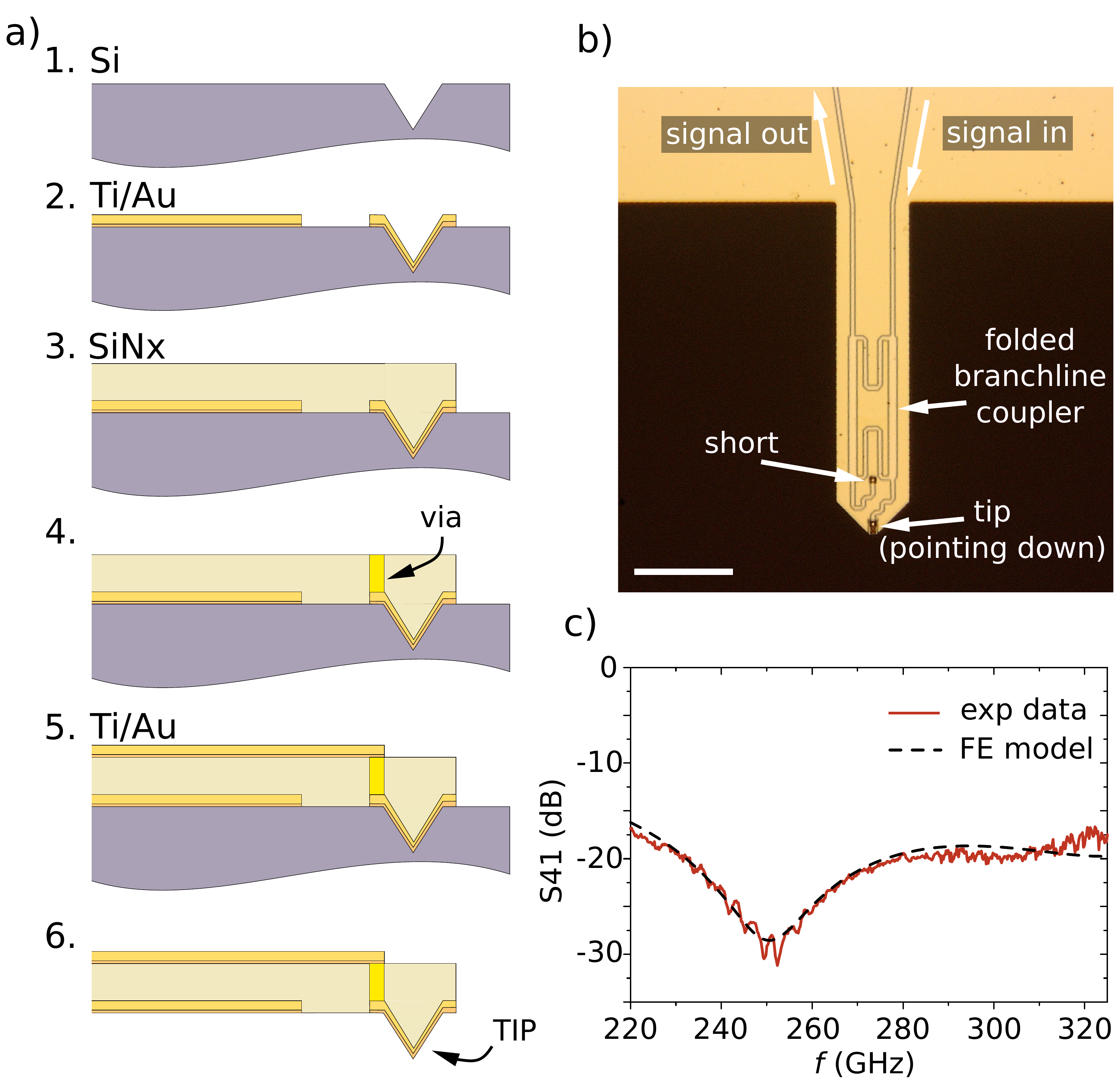}
 	\caption{\textbf{a)} Main fabrication steps of the THz cantilever. \textbf{b)} Optical image of the final, released cantilever. The scale bar corresponds to $75~\mu$m. \textit{Signal in} and \textit{signal out} denote the transmission lines connected to THz source and detector, respectively. The folded hybrid coupler patterned on the back of the cantilever is indicated. One arm of the coupler terminates in a short circuit, the other one terminates in the tip. \textbf{c)} Transmission amplitude from \textit{signal in} to \textit{signal out}. The transmission dip at 250 GHz indicates cancellation of the signal reflected from the open tip. The dashed curve shows the result obtained from finite element (FE) modeling.}
 	\label{fig:cantilever}
 \end{figure}

 We will now describe how this detection scheme can be implemented and used in a scanning probe cantilever to detect impedance changes at the probe tip.
 Figure \ref{fig:cantilever}b shows an optical microscope image of the shielded cantilever containing the THz circuit, patterned on its top side. The \textit{signal in} and \textit{signal out} lines (corresponding to ports 1 and 4 in Fig.~\ref{fig:branchline}a) are connected via landing probes with the source and detector (not visible). Since the dimensions of the cantilever (300 $\mu$m long, 75 $ \mu$m wide) are too small to host a circuit as shown in Fig.~\ref{fig:branchline}a, we have re-designed the branchline coupler such that the cross-branches are now folded inwards to fit the lateral dimensions of the cantilever. This slightly modifies the coupler properties. However, it does not change its basic functionality. 
% , whose cross branches are folded inwards to fit on the back of the 70 $\mu$m wide cantilever. This slightly changes the characteristics of the coupler, it does not, however, affect its basic functionality. 
 As discussed previously for the branchline coupler devices, one of the interferometer arms terminates in a \textit{short}. The other one, previously terminating in an \textit{open}, is now connected to the tip. We will keep the notation of the arms as introduced above, referring to the arm terminating in the tip as arm 2 with length $L_2$, and to the arm terminating in a short to ground as arm 3 with length $L_3$. In order to balance the coupler such that scattered signal cancellation is achieved, we have to take into account the finite capacitance of the open tip ($C_t \sim 2$ fF, obtained from finite element (FE) simulations) and adjust $L_3$ by $\Delta L$ accordingly. However, due to the folded geometry of the coupler and a resulting unwanted cross-coupling between the branches, significant leakage currents within the coupler result in a non-trivial relation between $\Delta L$ and signal cancellation at the detector line. Therefore, we use FE simulations to empirically determine a well-balanced configuration for the given $C_t$, for which we obtain $L_2 = 44~ \mu m$ and $L_3 = 54~ \mu m$, i.e. $\Delta L = 8~\mu$m . 
 
 \subsection{Fabrication}
In fig. \ref{fig:cantilever}a the fabrication flow for the cantilever is sketched.
In a first step (1) a pyramid shaped pit ($5~\mu$m deep) is etched into the Si wafer using KOH etching. This defines the position and shape of the tip. Next (2) we deposit (10+300)nm Ti/Au which serves as a ground plane. During this step also the pit is filled with a Ti/Au layer, which will become the metallic tip. The area around the pit is protected with an optical mask. The wafer is then (3) covered with 3 $\mu$m of PECVD SiNx that is subsequently etched with a Bosch process to define the geometry of the cantilever, $300~\mu$m long and $75~\mu$m wide. In a separate step $5\times 5~\mu$m$^2$ sized \textit{vias} are etched into the SiN$_x$ layer. These \textit{vias} will serve as an electrical connection between the MS top layer and the ground plane (to form a short) or with the cantilever tip, respectively.  
After a cleaning step, we pattern the strip lines with electron beam lithography and lift off techniques (5) and, in a separate step, connection of the\textit{ via} is established through angled deposition of Au. We use (2+300)nm Ti/Au bilayers. This step concludes the patterning of the transmission lines on the cantilever. Next, we release the cantilevers. In order to avoid exposure of the striplines to chemicals, we protect the surface of the wafer by gluing a Sp wafer on top of it with "black wax" (Apiezon W100).
We take particular care that no air bubbles remain in the wax to ensure a complete and efficient protection. The release step is prepared by patterning a SiN$_x$ mask on the backside of the wafer which contains windows at those positions where the cantilevers have been patterned on the front side of the wafer. The two wafers are then subjected to a KOH etch which etches through the windows on the backside of the wafer until the Ti/Au ground layer is reached. At this point the cantilever gets released from the Si wafer. Note, however, that on its front side it is still glued to the protection wafer. When the KOH etch is complete, the wafer is carefully immersed in Toluene to dissolve the black wax and to fully release the cantilever chips.  
Subsequently the cantilever is mounted on the landing probe setup. 
 
 \subsection{Experimental results}

The measured THz response of the cantilever is shown in Fig.\ref{fig:cantilever}c. We clearly observe a dip in transmission ($\sim -30$ dB) indicating a suppression of the scattered signal that gets reflected from the open tip into the detector line. We note that compared to the previously discussed branchline couplers without the tip (Fig.~\ref{fig:branchline}), the position of the dip is slightly shifted towards lower frequencies ($f = 250$ GHz). This is most likely a result of the folded geometry of the branchline coupler, consistent with a cross-capacitive coupling between neighbouring parts of the circuit. Moreover, the frequency shift as well as the relatively low transmission at higher and lower frequencies, suggest that the terminating capacitance of the tip slightly differs form the assumed value ($C_t = 2$ fF) such that the chosen $\Delta L = 8~\mu$m turns out to be not yet the best match. The FE model for the THz response (dashed line) yields good agreement with the experiment if we assume $C_t = 2.9~$fF and $R_{short} = 5~\Omega$. 

\begin{figure}
 	\includegraphics[width=1\linewidth]{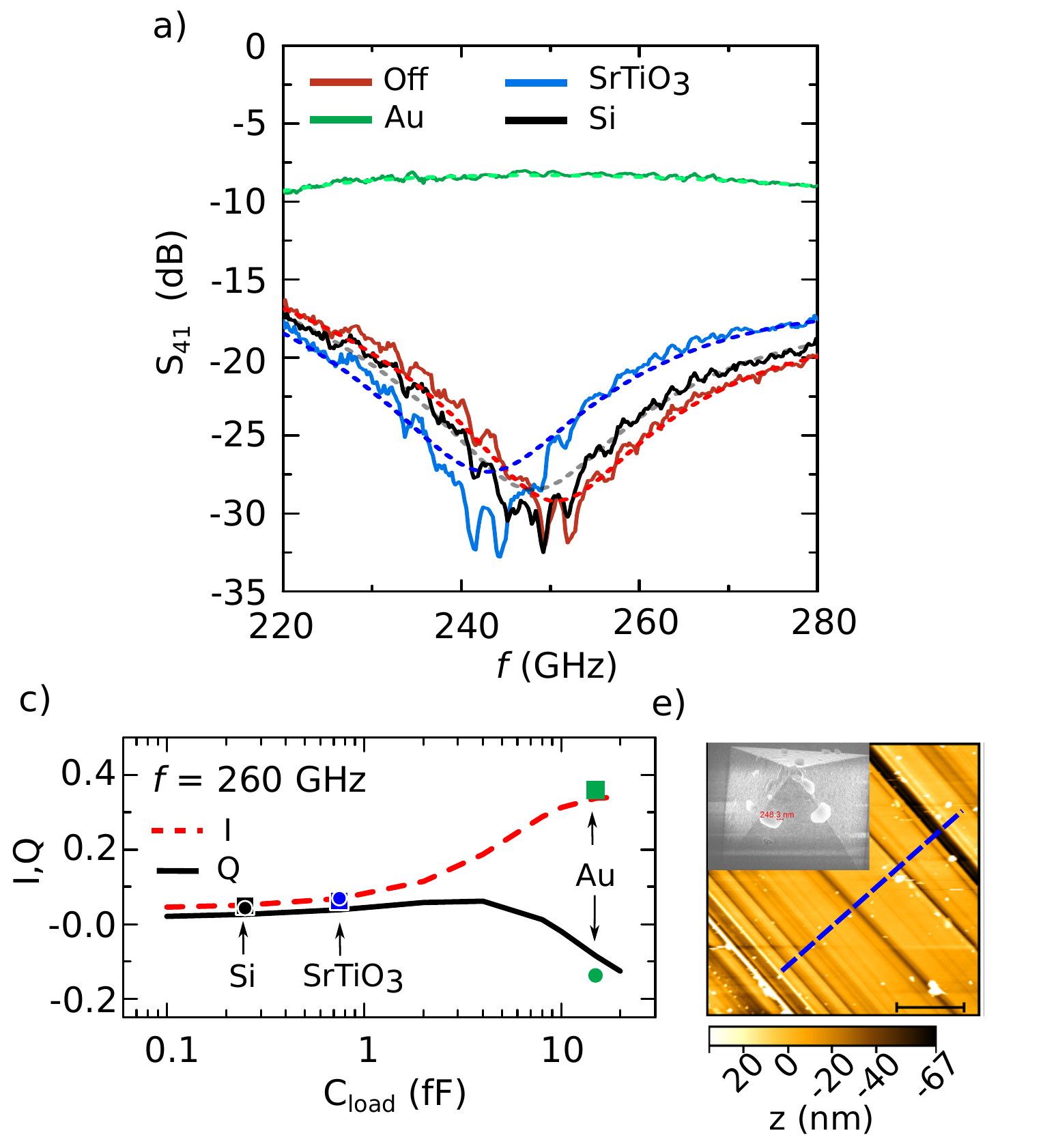}
 	\caption{ \textbf{a)} THz response of the cantilever when open (red) and landed on Si (black), SrTiO$_3$ (blue) and Au (green) samples. Solid lines correspond to measurements, dashed lines indicate finite element (FE) simulations with the following parameters: $R_\text{via} = 5 \Omega$, Tip capacitance: $C_t= 2.9~$fF; $L_2 =  44~\mu$m, $L_3 = 52~\mu$m, conductivity signal line and gnd plane: $\sigma_{au} = 19.2\times 10^{-6}$~S/m, $\epsilon_r(\text{SiN}_x) = 5.9$. $C_\text{load}$ (corresponding to $(1/C_{s,tip} + 1/C_{s,gnd})^{-1}$ in Fig. 1, $R_s = 0$) for Off: 0 fF; Si: 0.25fF; SrTiO$_3$: 0.75fF; Au: 15fF. \textbf{b)} I (dashed line, red) and Q (solid line, black) of the cantilever response at 260 GHz for different $C_{load}$ obtained from FE simulations. Symbols: experimental data extracted from a). \textbf{c)} $20~\mu$m$\times 20~\mu$m topography image of a scratched Si wafer, obtained with a cantilever similar to the one used in a) utilizing the beam deflection mode in a commercial AFM. Inset: SEM image of the cantilever tip. } 
 	\label{fig:TIM}
 \end{figure}

Our cantilever can be used to detect changes in the tip impedance when landed on a dielectric or metallic sample. This is demonstrated in Fig.~\ref{fig:TIM}. We have mounted the cantilever on the landing probe setup and we get the tip in contact with different materials, approached from below via a mechanical height control. Fig.~\ref{fig:TIM}a compares the measured response for the tip floating near the sample surface (red) and landed on 3 different materials, Au (green), Si (black) and SrTiO$_3$ (blue). 
%The corresponding optical images are shown in Fig.\ref{fig:TIM}b. 
We clearly observe distinct responses for each material. When brought into contact with a dielectric (Si,  $\epsilon_r = 11.9$ and SrTiO$_3$, $\epsilon_r = 300$) the dip shifts towards lower frequencies by $\Delta f_\text{(Si)} = 2$ GHz for Si and $\Delta f_\text{(STO)} = 10$ GHz for SrTiO$_3$. Notably, the overall line shape remains fairly similar. In contrast, upon contact with highly conductive Au ($\rho = 2\mu \Omega $cm), the dip vanishes and transmission is high over the full frequency range, as expected for a shorted tip.

\subsection{Discussion}

In order to quantitatively understand the cantilever response we use FE modelling of the full circuit and we include a load capacitance $C_{load}$ in parallel to $C_{t}$ to take into account contributions from the sample materials (cf. fig. \ref{fig:principle}a, neglecting Ohmic dissipation in the sample, $R_s = 0$). We find that the curves can be reproduced very well if we use $C_{Si} = 0.25 ~$fF, $C_{STO} = 0.75 ~$fF, and $C_{Au} = 15 ~$fF as the only adjustable parameter for each material. 
In Fig.~\ref{fig:TIM}b) we compare $I$ and $Q$ of the FE response for various $C_{load}$ with the experimental values obtained at $f = 260$ GHz for each sample. Since the response of the folded branchline coupler connected to the tip is slightly off resonance ($\sim 250$ GHz) we do not expect a simple linear behaviour as for the branchline coupler without the tip discussed previously. Figure \ref{fig:TIM}b shows that the response becomes more sensitive, i.e. the slope of the curves for $I$ and $Q$ becomes steep, for larger $C_{load}$ ($\sim 10$ fF). Since this is in the range of capacitance which we obtained for the Au sample, this indicates that our THz cantilever becomes more sensitive for metallic samples. In contrast to shielded microwave cantilevers, where sensitivity is highest around the metal-insulator transition \cite{lai2010mesoscopic}, for our cantilever the working point is shifted towards samples with higher conductivity. It may thus be used to detect electronic variations at high frequencies within a metal or even in superconductors in future scanning experiments.

Our THz cantilever is also compatible with AFM topography imaging. This is shown in Fig.\ref{fig:TIM} c) where a topography image of a scratched Si wafer surface is displayed, obtained using a THz cantilever mounted on a commercial Asylum Cypher AFM with a laser deflection read-out. Using a de-convoluting tip geometry modeling algorithm (\textit{Gwyddion} blind tip estimation algorithm \cite{nevcas2012gwyddion}) we estimate the tip apex to be $\approx$100 nm. An SEM image of a cantilever tip is shown in the inset in Fig. \ref{fig:TIM} c). 

Finally, we like to point out some aspects that we aim to improve for future THz cantilever generations. Firstly, even though our fabrication technique provides useful devices, the current yield is rather low ($\approx 10 \%$). This is mostly related to the use of the black wax, which is needed to protect the THz circuitry from chemicals during the release step, but which also induces mechanical stress, resulting in loss of a large number of cantilevers. Secondly, in its current design the substrate, which serves as a handling wafer, faces in the same direction as the tip. This limits the surface region on the sample, that can be reached by the cantilever to approximately the cantilever length ($\sim 300 ~\mu$m). In order to lift this constraint, developing a flip-chip technology may provide the most suitable means to bond a handling wafer to the top side of the cantilever chip. At the same time, however, it will be important to maintain access to the circuitry with landing probes. Thirdly, our experiments have shown that due to the folded geometry of the balanced branchline coupler the circuit response deviates slightly from that of the un-folded geometry tested on a substrate (Fig.~\ref{fig:cantilever}b in comparison to Fig.~\ref{fig:branchline}a). Therefore, FE simulations are required for a quantitative analysis of the measurement signal while a simple analytical equation would be more desirable. This may motivate a re-design of the cantilever such that it can host the balanced coupler without the need to modify its layout. In this case the measurements could be modeled within a textbook analytical description, which will strongly facilitate a quantitative interpretation of the measurement signal.

\section{Conclusion}
We have presented a shielded THz probe suitable for impedance microscopy in the sub-mm band (0.3 THz). As a key challenge for the realization of such a device we have identified the necessity to carry out common mode cancellation and impedance matching at THz frequencies close to the cantilever tip in order to enable sensitive detection of small changes of the tip impedance. We have addressed these challenges by developing an on-chip circuit that can be patterned on the cantilever which comprises a balanced branchline coupler. The coupler functions as an on-chip interferometer and in this manner achieves the required common mode suppression as well as high sensitivity to small impedance changes. To demonstrate the basic functionality of this concept, we have realized a set of devices on substrates and we have characterized their response at THz frequencies. The results can be directly modeled within an analytical model of the circuit. Furthermore, a fabrication technology has been developed that allows for patterning the circuit on a free standing cantilever including the tip. When the released cantilever is landed on different dielectric (Si, SrTiO$_3$) and metallic (Au) samples we observe distinct THz responses which enable us to determine the corresponding capacitive load at the cantilever tip using finite element modelling. Our cantilever removes several critical technological challenges towards scanning impedance microscopy at THz frequencies.         
 
\section*{Appendix: Analytical Model for the balanced branchline coupler}

In order to compute the response of the balanced hybrid coupler we describe the signal evolution in the coupler in terms of forward and backward travelling waves in the transmission lines and the reflection coefficents $\Gamma_\text{ms}$ and $\Gamma_\text{s}$ at the open and shorted transmission line, respectively. Using Kirchoff's rules for the voltage and current at each node of the hybrid, we can construct a system of equations that allows us to determine the voltage measured at the detector line at port 4 upon signal injection at port 1. 
%In order to compute the response of the balanced hybrid coupler, we construct an equivalent circuit that consists of two hybrid couplers that are connected in series such that port P2 and P3 of the first coupler (C1) connect with port P1 and P4 of the second one (C2). This is done to take into account that the signal reflected at the terminated arms gets re-injected into the coupler. The reflection coefficients at the termination of arms 2 and 3 are included as a transmission coefficient at the junctions P2$_\text{C1}$- P1$_\text{C2}$ and P3$_\text{C1}$- P4$_\text{C2}$. We can then decompose the network into a cascade of 2-port networks and use an odd-even mode approach to describe it in terms of a set of ABCD matrices \cite{pozar2009microwave}, relating voltages at each port to the respective impedance. We can then identify those coefficients relating the ingoing and outgoing voltage for zero current, which thus correspond to the scattering parameters.    
(The full set of equations is provided in the Supplementary material).
To describe wave propagation along each transmission line segment with length $L$ and impedance $Z$ we use the frequency dependent wave propagation factor $e^{\gamma L}$ and

\begin{align*}
&\gamma = \alpha \frac{Z}{Z_0} + i\beta\\
&\alpha = (1.1f \times 10^{-9} + 86.9)~\text{Np/m} \\
&\beta =  \frac{2 \pi}{c} f \sqrt{\epsilon_\text{eff}} 
\end{align*}

with $Z_0 = 55.5~\Omega$, $c = 3\times 10^8$ m(s)$^{-1}$, $\epsilon_{eff} = 4.47$  and $\alpha$ extracted from a direct measurement of a $130~\mu$m transmission line. For the low impedance lines we have used Z =Z$_l$ = 37 $\Omega$. The branch lengths of the coupler are $L = 130 \mu $m. 

To calculate the open terminating capacitance $C_t$ of the open MS line we have used 

\begin{align*}
C_{t} = G \frac{\sqrt{\epsilon_{eff}}}{c Z_0}, \\
G = \frac{\xi_1 \xi_3 \xi_5 h}{\xi_4}
\end{align*}

together with the following closed form expression:
\begin{align*}
&\xi_1 = 0.434907\frac{(\epsilon_{eff}^{0.81} + 0.26(w/h)^{0.8544} + 0.236)}{(\epsilon_{eff}^{0.81} - 0.189(w/h)^{0.8544} + 0.87)}\\
&\xi_2 = 1+\frac{(w/h)^{0.371}}{2.35\epsilon_r + 1}\\
&\xi_3 = 1 + \frac{0.5274~\tan^{-1}[0.084(w/h)^{1.9413/\xi_2}]}{\epsilon_{eff}^{0.9236}}\\
&\xi_4 = 1+ 0.037\tan^{-1}[0.067(w/h)^{1.456}] \\
&~~~~~~~~~~~~~~~~~~~~~~~~~~~~\times (6-5\exp(0.036(1-\epsilon_r)))\\
&\xi_5 =  1-0.218 \exp(-7.5(w/h))
\end{align*}

%\begin{equation*}
%\xi_1 = 0.434907\frac{(\epsilon_{eff}^{0.81} + 0.26(w/h)^{0.8544} + 0.236)}{(\epsilon_{eff}^{0.81} - 0.189(w/h)^{0.8544} + 0.87)}\\
%\end{equation*}
%\begin{equation*}
%\xi_2 = 1+\frac{(w/h)^{0.371}}{2.35\epsilon_r + 1}
%\end{equation*}
%\begin{equation*}
%\xi_3 = 1 + \frac{0.5274~\tan^{-1}[0.084(w/h)^{1.9413/\xi_2}]}{\epsilon_{eff}^{0.9236}}\\
%\end{equation*}
%\begin{align*}
%\xi_4 = 1+ 0.037\tan^{-1}[0.067*(w/h)^{1.456}] \\
%\times (6-5*\exp(0.036*(1-\epsilon_r)))
%\end{align*}
%\begin{equation*}
%\xi_5 =  1-0.218 \exp(-7.5(w/h))
%\end{equation*}

for which we have used the stripline width $w = 3.75~\mu$m, dielectric thickness $h = 3~\mu$m, $\epsilon_{eff} = 4.47$, PECVD SiN$_x$ dielectric constant $\epsilon_r = 5.9$ \cite{finkel2017performance}, characteristic impedance $Z_0 = 55.5~\Omega$ and the vacuum speed of light $ c=3\times10^{8}$ m/s.

\section{Acknowledgements}
We would like to thank Carmine de Martino and Luca Galatro for help with the THz characterization setup and David Thoen for support during the fabrication process development.
We further acknowledge scientific discussions with Andrea Neto, Nuria Llombart, Akira Endo, Jochem Baselmans, Ronald Hesper and Ivan Camara Mayorga.   
This research has been funded by the European Research Council Advanced Grant No.~339306 (METIQUM).

%\bibliography{Bib} 

\newpage

\mbox{}
\thispagestyle{empty}
\newpage

\appendix\onecolumngrid
\renewcommand{\thefigure}{S\arabic{figure}}
\setcounter{figure}{0}

\section*{Supplementary Material}
\subsection{Analytical model of the balanced hybrid coupler}

The circuit diagram of the hybrid coupler with the corresponding forward and backwards travelling wave voltage amplitudes $a_i$ is depicted in fig.\ref{fig:S1}. 
Using Kirchhoff's rules and Ohm's law we can derive the following condition for the sum of voltages at node 1:

\begin{figure}[b]
	\includegraphics[width=0.5\linewidth]{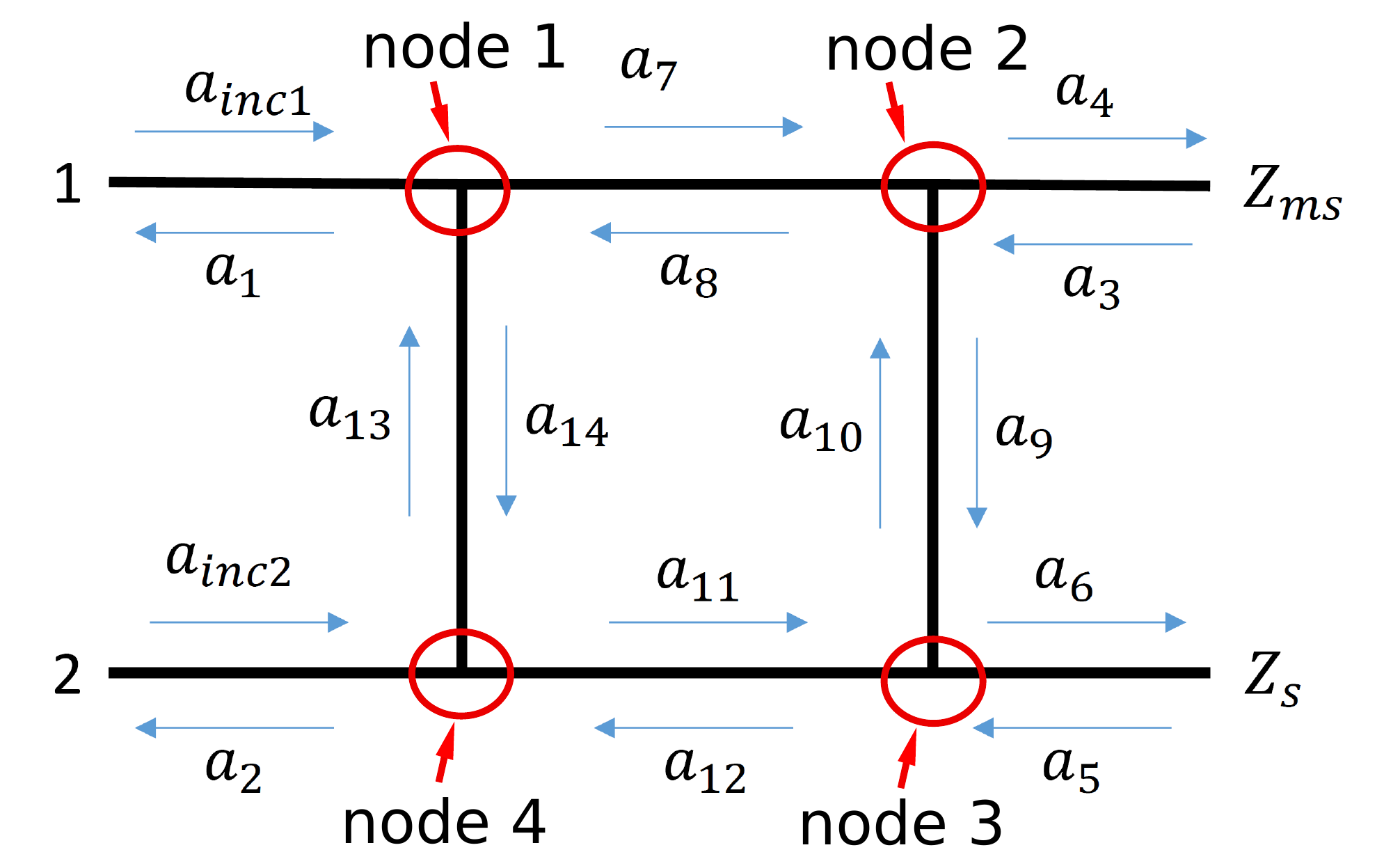}
	\caption{Circuit diagram for a hybrid coupler as described in the main text. $Z_{ms}$ and $Z_s$ denote the terminating complex impedance of the arms leading to the open stripline and to the short to ground, respectively.}
	\label{fig:S1}
\end{figure}

\begin{equation}
a_\text{inc1} + a_1 = a_7 + a_8 = a_{13} + a_{14},
\end{equation}
while we obtain for the currents at this point of the circuit
 \begin{equation}
 \frac{a_\text{inc1}}{Z_0} - \frac{a_1}{Z_0}-\frac{a_7}{Z_l}+\frac{a_8}{Z_l}+\frac{a_\text{13}}{Z_h}-\frac{a_\text{14}}{Z_h} = 0.
 \end{equation}

Likewise, we obtain for node 2
\begin{equation}
a_3 + a_4 = a_7F_l + a_8B_l = a_{9} + a_{10},
\end{equation}

and

\begin{equation}
 \frac{a_3}{Z_0} - \frac{a_4}{Z_0}+\frac{a_7F_l}{Z_l}-\frac{a_8B_l}{Z_l}-\frac{a_\text{9}}{Z_h}+\frac{a_\text{10}}{Z_h} = 0.
\end{equation}

For the arm to the open microstrip terminating with $Z_{ms}$ we get

\begin{equation}
- \frac{a_3B_{ms}}{Z_0} + \frac{a_4F_{ms}}{Z_0} = \frac{a_3B_{ms} + a_4F_{ms}}{Z_{ms}}
\end{equation}

For nodes 3 and 4 we obtain in the same fashion:

\begin{align}
&a_6 + a_5 = a_{11}F_l + a_{12}B_l = a_{9}F_h + a_{10}B_h,\\
&0 = \frac{a_5}{Z_0} - \frac{a_6}{Z_0}+\frac{a_{11}F_l}{Z_l}-\frac{a_{12}B_l}{Z_l}+\frac{a_\text{9}F_h}{Z_h}-\frac{a_\text{10}B_h}{Z_h},\\
&- \frac{a_5F_s}{Z_0} + \frac{a_6B_s}{Z_0} = \frac{a_5B_s + a_6F_s}{Z_{s}},\\
&a_\text{inc2} + a_2 = a_{11}F_l + a_{12}B_l = a_{13}B_h + a_{14}F_h,\\
&0 = \frac{a_\text{inc2}}{Z_0} - \frac{a_2}{Z_0}-\frac{a_{11}F_l}{Z_l}+\frac{a_{12}B_l}{Z_l}-\frac{a_\text{13}B_h}{Z_h}+\frac{a_\text{14}F_h}{Z_h},
\end{align}

This directly leads to the set of equations

\begin{align}
-a_{inc1} &= a_1 - a_7 - a_8  \\
0 &= a_7 + a_8 - a_{13} - a_{14} \\
-a_\text{inc1} &= -a_1 - a_7 \frac{Z_0}{Z_l} + a_8\frac{Z_0}{Z_h} + a_{13}\frac{Z_0}{Z_h} - a_{14}\frac{Z_0}{Z_h}\\ 
0 &= -a_3 - a_4 + a_7F_l + a_8B_l \\
0 &= a_3 + a_4 -a_9 -a_{10}\\
0 &= a_3 - a_4 + a_{7}F_l\frac{Z_0}{Z_l} -a_{8}B_l\frac{Z_0}{Z_l} - a_{9}\frac{Z_0}{Z_h} + a_{10}\frac{Z_0}{Z_h}\\
0 &= a_3 B_{ms} (\frac{Z_0}{Z_{ms}} + 1) + a_4 F_{ms}(\frac{Z_0}{Z_{ms}}-1) \\
0& =-a_5  -a_6 + a_9F_h + a_{10}B_h \\
0 &= a_5 + a_6  -a_{11} -a_{12}  \\
0 &= a_5 -a_6 + a_9 F_h \frac{Z_0}{Z_h}  -a_{10} B_h \frac{Z_0}{Z_h}  -a_{11}\frac{Z_0}{Z_l} + a_{12}\frac{Z_0}{Z_l}  \\
0&= a_5 B_{s}(\frac{Z_0}{Z_{s}}+1) + a_6 F_{s}(\frac{Z_0}{Z_{s}}-1) \\
0&= -a_{11}F_l -a_{12}B_l + a_{13}B_h + a_{14}F_h \\
a_{inc2} &= -a_2 + a_{11}F_l + a_{12}B_l  \\
a_{inc2} &= -a_2 + a_{11}F_l \frac{Z_0}{Z_l}  -a_{12}B_l \frac{Z_0}{Z_l}  -a_{13}B_h \frac{Z_0}{Z_h} +a_{14}F_h \frac{Z_0}{Z_h}
\end{align}

with 
\begin{align*}
&F_l = \exp(-\gamma_l L),~~~~~~~~~~~
&\gamma_l = \alpha \frac{Z_l}{Z_0} + i\beta ,~~~~~~~~~~~
&F_h = \exp(\gamma_h L) ,~~~~~~~~~~~
&\gamma_h = \alpha \frac{Z_h}{Z_0} + i\beta ,~~~~~~~~~~~
&B_l = \exp(-\gamma_l L),~~~~~~~~~~~\\
&B_h = \exp(\gamma_h L),~~~~~~~~~~~
&B_{ms} =  \exp(\gamma L_2) ,~~~~~~~~~~~
&F_{ms} = \exp(-\gamma L_2),~~~~~~~~~~~
&\gamma = \alpha + i\beta,~~~~~~~~~~~
&B_{s} = \exp(\gamma L_3),~~~~~~~~~~~ \\
&F_{s} = \exp(-\gamma L_3)
%&P = \frac{Z_0}{Z_h}\\
%&Q = \frac{Z_0}{Z_l}\\
%&F_l = \exp(-\gamma_L l)\\
%&F_h = \exp(\gamma_h l) \\
%&B_l = \exp(-\gamma_l l)\\
%&B_h = \exp(\gamma_h l)\\
%&R =  \\
%&Ft=\exp(-\gamma ToTip); \\
%&Bt=exp(gammaNot(i)*ToTip); \\
%&Fs=exp(-gammaNot(i)*ToShort); \\
%&Bs=exp(gammaNot(i)*ToShort); \\
%&Tip1 = B_{tip} (\frac{Z_0}{Z_{TIP}} + 1)\\
%&Tip2 = F_{tip}(\frac{Z_0}{Z_{TIP}}-1) \\
%&Short1 =  B_s(\frac{Z0}{Zs}+1) \\
%&Short2 = F_s(\frac{Z_0}{Z_s}-1)
\end{align*}
$\alpha$ and $\beta$ are as given in the appendix to the main text. $L = 130~ \mu$m is the branch length of the coupler, $L_2 = 49~\mu$m the arm length to the open termination and $L_3 = L_2 + \Delta L$ the length of the arm terminating in a short to ground, as given in the main text. The terminating impedance of the open microtrip line is frequency dependent, and given by $Z_{ms} = f/(i C_t 2\pi) $ with $C_t = 0.163$~fF (see appendix of the main text). The impedance of the short-to-ground terminated arm is purely Ohmic, $Z_s = 1.6~\Omega$.

To obtain the response for signal incident at port 1 we set $a_{inc1} = 1$ and $a_{inc2} = 0$ and solve the set of equations A7 - A20 at each frequency. Coefficient $a_2$ (cf. fig.~\ref{fig:S1}) then corresponds to the signal measured in the detector line, as plotted in the main text.

\newpage
\subsection{Detailed fabrication procedure of the THz cantilever}

\begin{enumerate}
	\item Etching of substrate to define the tip position
	\begin{itemize}
		\item \textit{Preparation of the wafer}. Take 4” HRFZ Si wafer with 50nm LPCVD SiN. Spin PMMA 950K A6 at 3000 rpm speed. Bake for 2 min @ 180\textcelsius~on the hot plate.
		\item \textit{E-beam lithography}. We use a Raith EBPG 5200. Align the center in the aligning microscope, adjust the rotation so that the flat side is along microscope’s Y axis (horizontal). 
		Dose 1250 $\mu$C/cm2, highest beam current with spot size >100nm, step size 70nm. Develop in MIBK:IPA 1:1 standard solution for 60s, dry in spin dryer for 2 min, check pattern in the microscope.
		\item \textit{Dry etch}. We use a Leybold Fluorine RIE. Dry etch for ~60s in CHF3 with end point detection laser focused on the open square in the center. 
		\item Strip the PMMA in Acetone overnight + fuming nitric acid for 1 hr. Check the pattern in the microscope.
		\item \textit{Prepare the KOH 30 \% bath}. Heat up KOH bath to 80\textcelsius. After the bath is ready, 10-15s of CHF3 or SF6 plasma etch to remove native oxide. Immediately after (<5min) start the KOH etching. Put wafer in the bath using plastic holder, wait ~ 6 min, rinse in DI water several times. Check the pattern in the microscope. When defocused, pyramids should show a clear diagonal cross without spot in the center. If there is doubt, add extra 2 mins in KOH bath followed by careful DI rinsing.
		\item \textit{Etching SiN on the front side}. Spin AZ40XT photoresist @ 2000 rpm on the back side (where there is no pattern yet) for protection against HF. Pay attention that the chuck is clean. Bake carefully on a hot plate starting at 60 \textcelsius, increasing temperature by 5 \textcelsius~every 5 min, up until 100\textcelsius. Bake at 100\textcelsius~ for extra 15 min. Make sure the hot plate is clean before starting the baking, or use foil. After the resist on the backside is baked, etch in 40\% HF for 15 min until tip side turns hydrophobic. Rinse carefully, dry and check in the microscope. The lighter boundary of undercut SiN around the tip square now should be gone. Remove AZ40XT in acetone overnight + rinse in fresh acetone 10 m + 1 hr fuming nitric acid.
		\end{itemize}
	\item Ground layer
	\begin{itemize}
		\item \textit{optical lithography.} Make sure the wafer is clean. Spin AZnLOF 2020 at 3000 rpm, bake at 100 C hot plate for 2 min. Expose the wafer in EVG setup using TIPS layer markers (the only present so far), make sure the flat on the mask corresponds to the flat on the wafer, also in aligning microscope the orientation of the labels on mask and wafer should match. Contact exposure for 15 s, 2 min post exposure bake at 110 C hot plate, develop in AZ826 MIF developer for 60 s (10 s after clearing of unexposed areas), rinse in DI water, dry, check pattern in the microscope. Critical feature is the gap around the tip and the break-off bridge from tip to the bulk. When defocusing at tip location the same clear diagonal cross should show up, if not really clear add some extra time in the developer.
		\item \textit{Cleaning step.} 10-30 s of O2 plasma etching, (Leybold Fluor), 50 W, 50 mBar, 50 sccm. It is very important there are no resist residues in clear areas. If there is unremoveable dirt better clean thoroughly and start AZnLOF over.
		\item \textit{Deposition}. Deposit 10 nm Ti + 300 nm Au in Temescal at normal angle, rates are 0.5 A/s for Ti and 1 A/s for Au. At higher rates the heat will melt photoresist causing troubles in lift-off.
		\item \textit{Lift-off}. Overnight in NI 555 resist remover from AZ at 80 \textcelsius, also in ultrasonic bath at 60\textcelsius. Flush with DI water shower at full strength. Rinse in IPA, water, 1 hr fuming nitric acid. ANY organic residue can spoil next PECVD SiN deposition step, and that is irrecoverable.
	\end{itemize}
	\item Deposition of the SiNx dielectric
	\begin{itemize}
		\item Make sure wafer is clean from any organics!
		Deposit 3 um layer of (high quality) PECVD SiN (Oxford Instruments PlasmaPro 80) 6 hrs.
	\end{itemize}
	\item Dry Ecth of the \textit{vias}
	\begin{itemize}
		\item \textit{Deposit Cr-layer}. Deposit 100 nm Cr (Temescal FC-2000).
		\item \textit{Spin resist} Spin PMMA 950K A6 at 3000 rpm, bake for 5 min on 150\textcelsius hotplate.
		\item \textit{E-beam lithography of the Cr hard mask}. Use ground layer markers for auto alignment, same recipe as e-beam in step 1. Develop in IPA:MIBK 1:1 for 60 s, check in the microscope.
		\item \textit{cleaning step.} 10-30 s of O2 plasma in F1 or F2, 50 W, 50 mBar, 50 sccm. It is very important there are no resist residues in clear areas.
		\item \textit{etch Cr hard mask}. Cr layer etch in Cr 01 (from Microchemicals) wet etchant for 40 s (10 s after open features become dark), rinse carefully in DI water, dry, check in the microscope that the smallest features are open and the largest are uniformly open.
		\item When the Cr mask is of sufficient quality, strip PMMA in acetone/nitric acid.
		\item Check in the microscope that all uncovered areas in the mask are uniform.
		\item Stick the wafer to 1mm Si carrier wafer using organic white thermal paste. Apply paste uniformly leaving ~5 mm clear area near the edge, press the wafer to the carrier wafer with paste on it on a hard surface with cleanroom tissue over.
		\item \textit{Dry etch} SiN in Bosch etcher, usually takes 7-7.5 minutes, wait 1 min after the EPD level saturates. EPD must be focused on the open square in the center.
		\item Check the resistivity of contact area of uncovered gold near the flat, both in 4W and 2W modes using 4W probe, if resistivity is OK check in the microscope the smaller features and gold color there, should be clean gold without change in color between different places on the wafer.
		\item Dissolve white paste in IPA/acetone (ultrasonic bath 40\textcelsius), if necessary wipe white paste residues with IPA cleanroom tissue, rinse in IPA 5min, rinse in DI water, dry up.
		\item Strip Cr layer in Cr01 (30 min).
	\end{itemize}
	\item Filling of the \textit{vias}
	(here we descripe the procedure using electroplating)
	\begin{itemize}
		\item Fix the wafer in the holder of the electroplating setup with a bronze spring over an uncovered Au area near flat edge of the wafer, connect the spring to the WE wire using a crocodile clamp. Put the wafer in the solution so that the open areas near the flat edge do not touch the solution. Gold plate for 10 min with frequent checks in a profiler (dektak), rinse carefully, dry and check in the microscope that vias look filled up (no significant focus shift between surface in the via and SiN surface, larger features like ground landing pads have slightly rougher surface).
		\item When sufficiently plated, clean the wafer in DI water excessively, check in optical microscope, possibly in AFM.
	\end{itemize}
	\item Dry etching of the cantilevers
	\begin{itemize}
		\item Same recipe as step 4, Cr mask, e-beam, O2 plasma, Bosch. Or SF6 plasma.
	\end{itemize}
	\item Deposition of the stripline layers
	\begin{itemize}
		\item Deposit 30 nm Cr (Temescal FC2000). Spin PMMA 950K A9, 3000 rpm, bake 15 min at 150 C.
		\item \textit{E-beam lithography} as in step 4, 1250 dose, develop 1:1 60s, check in microscope, O2 plasma, Cr development in Cr01, check in microscope.
		\item Mount in Temescal, pump overnight, ion gun sputtering/cleaning for 30 s, deposit Ti-Au (2nm-300nm). Lift-off in NMP (ultrasonic bath, 60\textcelsius~, overnight). Nitric acid clean 15 min. Rinse in H$_2$O. Check in the microscope. Etch Cr conducting layer in Cr01 (15min). Possibly check \textit{vias} in probe station using central chip test structures.
	\end{itemize}
	\item Backside windows
	\begin{itemize}
		\item For protection of the front side use a layer of regular photoresist. Spin regular photoresist on the backside at 3000 rpm, bake 2 min @ 100\textcelsius~ on the hotplate. Use backside alignment in mask aligner (EVG 620 NUV), 6s exposure, developer 60 s. Check in microscope, start over if necessary.
		\item Dry etch SiN on the backside in SF6 or CHF3 plasma. Strip photoresist (acetone+nitric acid).
	\end{itemize}
	\item Black wax bonding
	\begin{itemize}
		\item Put the wafer on a hotplate covered with Al foil, heat up to 120\textcelsius~, paint it’s circuitry side carefully with black wax stick (Apiezon W100) leaving ~8 mm space near the edge uncovered. After cooling down, put the wafer in the bonding tool (EVG520 Wafer bonder), put flags on top, put Sp wafer on top of flags, align Sp to the wafer. Close the lid of the bonder, pump down to ~1mbar, heat up to 120\textcelsius, apply center pin, wait 30 min, apply min pressure, wait 2 hrs or more, remove pressure, cool down, vent). After that the black wax should be uniformly distributed between wafers. Clean the bonding tool table carefully with toluene, wipe with IPA.
	\end{itemize}
	\item Removal of the substrate/Cantilever release
	\begin{itemize}
		\item \textit{KOH bath}. After the temperature reached 80\textcelsius, etch native oxide on the wafer with 10 s CHF3 plasma etch, immediately after that start KOH etching.
		\item After 4 hrs of etching at 85\textcelsius, we move the wafer stack to a special KOH bath with horizontal positioning, keep etching at 70\textcelsius~for 6 hrs (until the membrane/gold is visible). Rinse carefully in water, then in IPA, let it dry vertically, inspect in microscope. Soak in DI water several times, a few hrs in total. Dissolve black wax in Toluene, soak the wafer separately in several fresh toluene solutions until perfectly clear (~1 day in total), rinse sereval times in PA, let it dry vertically. Break the wafer into chips.
	\end{itemize}
\end{enumerate}

\end{document}